\documentclass[aps,pra,reprint,10pt,a4paper,nofootinbib,showpacs,varwidth]{revtex4-1}
\usepackage[utf8]{inputenc}
\usepackage{graphicx, graphics}
\usepackage{newlfont}
\usepackage{epsfig,amssymb,times,amsfonts,amsmath,amsthm,braket,color,xcolor}
\usepackage{mathbbol}
\usepackage{bm}
\usepackage[normalem]{ulem}
\newcommand{\stkout}[1]{\ifmmode\text{\sout{\ensuremath{#1}}}\else\sout{#1}\fi}

\begin{document}
	


\title{Estimating quantum coherence by noncommutativity of any observable and its incoherent part}

\author{Tanaya Ray, Ahana Ghoshal, Arun Kumar Pati, Ujjwal Sen}

\affiliation{Harish-Chandra Research Institute, HBNI, Chhatnag Road, Jhunsi, Allahabad 211 019, India}

\begin{abstract}
	We establish 
	a lower bound on the 
quantum coherence of an arbitrary quantum state in arbitrary dimension, using a noncommutativity estimator of an arbitrary observable of sub-unit norm, where the  estimator is the commutator of the observable and its incoherent or classical part.
	The relation provides a direct method of obtaining an 
	estimate of the quantum coherence of an arbitrary quantum state, without resorting to quantum state tomography or the existing witness operators.
\end{abstract}

\maketitle

\section{Introduction}

The concept of quantum coherence and the uncertainty relations form two fundamental pillars of quantum mechanics.  They give rise to phenomena and applications in quantum systems that are significantly different from those in classical ones, and distinguish the two worlds in a quantitative manner. The linearity in the quantum description of physical states leads to the possibility of superposition of states of a quantum system, and this leads to the existence of quantum coherence. While the concept of quantum coherence was known since the early days of quantum theory, it is only recently that it has been provided with a careful quantification 
and a resource-theoretic analysis
\cite{eiTa-coherence}. It has since been 
widely useful in quantum technologies like quantum-enhanced metrology \cite{eiTa-metro},
quantum algorithms \cite{eiTa-algo, egulo-alada-2}, and quantum state discrimination \cite{eiTa-discrimi}.
Quantum coherence has been also found useful in the diverse fields of thermodynamics \cite{eiTa-thermo}, and has been argued to be functional in certain branches of biology \cite{eiTa-bio} as well. 

Noncommutativity between two or more observables, ultimately also related to the superposition principle, is another fundamental feature of quantum mechanics. Quantum uncertainty relations \cite{eiTa-uncertainty} quantify the noncommuting nature of quantum observables, and are useful for a wide range of applications that span from the foundations of physics all the way to technological applications. 
In particular, the uncertainty relations have been utilized
for 
entanglement detection \cite{eiTa-uncertainty-ent} and  for  security analysis in quantum key distribution \cite{eiTa-uncertainty-qkd}.

Quantum coherence can be quantified through a variety of approaches. A typical measure begins with the identification of a preferred orthogonal basis (``pointer states''), the probabilistic mixtures of  states of which are deemed as ``incoherent'' states. (See \cite{egulo-alada, egulo-alada-2}, however.) The failure to fall in that class of incoherent states is quantified in several ways, one of which is to ``accumulate''
the off-diagonal terms of the quantum density matrix expressed in the preferred basis. When the  \(l_1\)-norm is used to perform the accumulation, the corresponding measure is referred to as the \(l_1\)-norm of quantum coherence \cite{eiTa-coherence}. This is the measure that we will use to quantify the notion of quantum coherence.





The quantum uncertainty relations tell us that the noncompatibility between two observables of a quantum system can be quantified by the average value of their commutator. We will estimate the noncompatibility between an arbitrary observable and its ``incoherent part'' by using the average value of their commutator. The incoherent part of the observable is created by stripping out the off-diagonal parts of the observable when expressed in the basis that was chosen to be preferred for estimating quantum coherence. 


In this paper, we establish a relation  between the quantum coherence of an arbitrary quantum state, pure or mixed, in a quantum system of arbitrary dimension and the noncompatibility estimate of an arbitrary observable of sub-unit norm of that system.
Along with being potentially of fundamental use, it also provides a ready estimate---precisely, lower bound---on the quantum coherence of an arbitrary quantum state, without recourse to quantum state tomography. 
Indeed, every choice of an observable with sub-unit norm provides a potentially independent 
lower bound. As a by-product, it provides a direct method of creating witnesses for quantum coherence, unrelated 
to the existing witness operators.



The remaining part of the paper is arranged as follows. In Sec. \ref{cal}, we provide formal definitions of the physical quantities that we use, and a proof of the relation for arbitrary  states in any dimension. A short separate proof is provided for the qubit case.
%
%
%
%
A proof of the relation for the general case---arbitrary quantum states, pure or mixed, of an arbitrary dimension---is given in Sec.~\ref{pudim}. 
In Sec.~\ref{chharpokader-khulechhe-borat}, we consider an alternative definition of quantum coherence, using the concept of convex roof, and again find that the same relation is true even for this altered definition. 
A discussion is presented in Sec.~\ref{phurtir-pran-gaRer-maTh}.




\section{Uncertainty commutator \& quantum coherence}
\label{cal}

\noindent \textbf{Quantum coherence.--} Quantum coherence is naturally a basis-dependent concept, which is why we first need to fix the preferred, or reference, basis. Given a $d$-dimensional  Hilbert space  $ \mathbb{C}^d$ ($d$  is  assumed finite), we will assume that the physical set-up of the corresponding quantum  system dictates that its reference orthonormal basis  be \( \{\ket{j}\}\), \(j= 1,2, \ldots, d\). The density matrices that are diagonal in this specific basis are considered to represent the incoherent states of the system. Hence, all incoherent density operators $\rho^{\mathcal{I}} \in \mathcal{I}$, where \(\mathcal{I}\) denotes the set of all incoherent states for the reference basis considered, are of the form 
\begin{equation}
\rho^{\mathcal{I}}=\sum_{j=1}^{d} p_i \ket{j}\bra{j},
\end{equation}
with \(\{p_i\}\) forming a probability distribution.
Any state described by a density matrix outside $\mathcal{I}$ is a quantum coherent state. The off-diagonal elements in the density matrix, when expressed in the reference basis, give rise to the quantum coherence in the state, and their quantification is attained by using a suitable function of the same.
In this paper, we use the intuitive ``$l_1$ norm of quantum coherence''  to quantify 
the quantum coherence content in a quantum state. (See \cite{eiTa-coherence} in this respect.) More precisely, 
we define the quantum coherence of an arbitrary \(d\)-dimensional state as the sum of the moduli of the 
off-diagonal terms when the state is expressed in the reference basis. Therefore, for an arbitrary
\(d\)-dimensional state
\begin{equation}
   \rho=\sum_{i,j} p_{ij}\ket{i}\bra{j},
\end{equation}
expressed 
in the reference basis,
the 
quantum coherence of $\rho$ is given by  
\begin{equation}
\label{footboard-jhule-jawa}
C(\rho)= \sum\limits_{i\neq j}|p_{ij}|.
\end{equation}  
Here, 
\(p_{ij}\)'s are arbitrary complex numbers, with \(\sum_i p_{ii} = 1\).

There are measures of quantum coherence that are different from the  $l_1$ norm of quantum coherence, and they are not 
all equivalent. In particular, for any measure of quantum coherence, $C$, one can define the convex roof of \(C\), 
similar to what is done, e.g., for defining the entanglement of formation~\cite{eiTa-ent-of-formation, woh-achanak-a-gayi}. The convex roof of \(C\)
is defined as 
\begin{equation}
 \label{neche-neche-ai-ma-shyama}
\tilde{C}(\rho) = \inf \sum p_i C(|\psi_i\rangle) ,
 \end{equation}
 for an arbitrary density matrix $\rho$ on \(\mathbb{C}^d\), where the infimum is over all decompositions of \(\rho\) into \(\sum_ip_i|\psi_i\rangle \langle \psi_i|\). See e.g., Refs. \cite{Yuan, Winter} in this regard.
 Of course, we have $\tilde{C}(\rho)=C(\rho)$ if  $\rho$ is pure, and $\tilde{C}(\rho)=C(\rho)=0$ if $\rho$ is an incoherent state.
In~\cite{Xu}, they have  established that the convex roof  of the $l_1$ norm of quantum coherence is not equal to the $l_1$ norm of quantum coherence, for a mixed state on \(\mathbb{C}^3\).
\\

\noindent \textbf{Uncertainty commutator.--} The ``Heisenberg-Robertson'' quantum uncertainty relation bounds the product of the variances of two observables by using 
the expectation value of their commutator \cite{eiTa-uncertainty}.
The average value of the commutator between two observables, therefore, provides an estimate of the 
incompatibility of the two observables. We will here consider the same estimate of incompatibility between an observable and its ``incoherent part'' (defined below) to bound the quantum coherence of the relevant state.

Let us consider an observable $A$ on the Hilbert space \(\mathbb{C}^d\), so that \(A\) is  a Hermitian operator on \(\mathbb{C}^d\), and a state of the system under consideration, represented by the density matrix $\rho$. 
Then the uncertainty in measuring $A$ can be represented by
\begin{equation}
\bigtriangleup^2 A_\rho = \mbox{tr} (A^2 \rho)-(\mbox{tr} (A\rho))^2 .
\end{equation}

Consider now the diagonal part of the state $\rho$, and refer to it as \(\rho^D\), so that  
\begin{equation}
\rho^D =\sum \limits_{j} \bra{j}\rho\ket{j} \ket{j}\bra{j} .
\end{equation}
This matrix is just the density matrix \(\rho\), with only the diagonal terms, when written in the 
reference basis \(\{|j\rangle\}\), i.e., it is the incoherent part of \(\rho\), with respect to the reference basis.
We can similarly write down the diagonal or the incoherent part of \(A\), and refer to it as \(A^D\), with respect to the reference basis. 
The incoherent part, \(A^D\), of \(A\), is given by 
\begin{equation}
A^D=\sum\limits_{j} \bra{j} A \ket{j} \ket{j}\bra{j} .
\end{equation}
The observable $A$ can then be expressed   in the computational basis as
\begin{equation}
A=\sum\limits_{k,j} \ket{k} \bra{k} A  \ket{j}\bra{j}=A^D+ \sum\limits_{k \neq j}  A_{kj}  \ket{k}\bra{j} .
\end{equation}

Consider now the commutator between the observable \(A\) and its incoherent part, and let us compute its expectation in the
state $\rho$. This expectation value is given by
\begin{equation}
\label{ekhane-amar-maTidesh}
\frac{1}{2} \text{tr}\big([A,A^D]\rho\big) 
=i\text{Im}\sum_{i,j}p_{ij}A_{ji}A_{ii},
\end{equation}
where ``Im'' stands for the imaginary part of its argument.
The factor of \(1/2\) is taken in hindsight, as will become clear later. 
\subsection{Case of an arbitrary qubit}
Consider  a general  single-qubit state, 
$\rho=\frac{1}{2}(I+\overrightarrow{v}\cdot\overrightarrow{\sigma})$, where $I_2$ is the identity operator on the qubit Hilbert space, $\overrightarrow{v}=(v_x,v_y,v_z)$ is a real vector (Bloch vector) satisfying $v_x^2+v_y^2+v_z^2 \le 1$, and \(\overrightarrow{\sigma} = (\sigma_x, \sigma_y, \sigma_z)\) is the vector of quantum spin-1/2 Pauli matrices. 
The operator \(A\) is now a Hermitian operator on the qubit Hilbert space, and restricting to traceless operators with unit norm, we have 
\(A=\overrightarrow{\sigma} \cdot\widehat{n}\)
where \(\widehat{n} = (n_x, n_y, n_z)\) is a real three-dimensional unit vector.
Equation (\ref{ekhane-amar-maTidesh}) can now be rewritten as
\begin{equation}
\frac{1}{2} \text{tr}\big([A,A^D]\rho\big)
=-in_z\big(v_xn_y+v_yn_x\big).
\end{equation}
Now, $|v_xn_y+v_yn_x|=|(v_x,v_y)\cdot(n_y,n_x)| \le \sqrt{(v_x^2+v_y^2)(n_x^2+n_y^2)} \le \sqrt{v_x^2+v_y^2}$, where we have used the Cauchy-Schwartz inequality. Therefore, for an arbitrary single-qubit quantum state, we have the inequality
\begin{equation}
\frac{1}{2} \Big|\text{tr}\big([A,A^D]\rho\big)\Big| \le \sqrt{v_x^2+v_y^2}.  \end{equation}
On the other hand, 
for such a single qubit state, the \( l_1 \)-norm of quantum coherence in the \(\{\ket{0},\ket{1}\}\) basis 
is equal to $\sqrt{v_x^2+v_y^2}$.
So, for an arbitrary single qubit case, we have the relation
\begin{equation}
\label{ektara}
  \frac{1}{2} \Big|\text{tr}\big([A,A^D]\rho\big)\Big| \le C(\rho).
\end{equation}
The quantum coherence of a single qubit, therefore, is related to the ``noncommutativity'' or the ``uncertainty commutator'' of an arbitrary observable and its incoherent part, providing thereby a method of estimating the former physical quantity.

\subsection{Higher dimensions}
\label{pudim}
Let us now consider $A$ to be a traceless Hermitian operator on the $d$-dimensional complex Hilbert space.
The space of Hermitian operators on \(\mathbb{C}^d\) can be spanned by a basis of the form  
$\{I,\{\mathbb{H}_\mu\}_{\mu=1}^{d^2-1}\}$,
where $\mathbb{H}_\mu$'s are 
traceless Hermitian operators on $\mathbb{C}^d$, and where \(I\) denotes the identity operator on the same space. 
We further assume that 
${\mathbb{H}_\mu}$'s are orthonormal, i.e., \( \mbox{tr}({\mathbb{H}_\mu}^\dagger {\mathbb{H}_\nu} )= \delta_{\mu,\nu} \). We are therefore considering the inner product in the space of operators on \(\mathbb{C}^d\) as \(\mbox{tr}(\mathbb{A}^\dagger \mathbb{B})\) for two arbitrary operators \(\mathbb{A}\) and \(\mathbb{B}\) on \(\mathbb{C}^d\).
We can therefore write the traceless Hermitian operator $A$  as 
\(
A=\sum_{\mu=1}^{d^2-1}a_\mu \mathbb{H}_\mu
\)
where  $a_i \in \mathbb{R}$.

Let us now consider the traceless Hermitian operators with unit norm only:
 $\sum\limits_{\mu=1}^{d^2-1}  a_\mu ^2= 1 $.  
 The norm of the operators is of course defined via the above-mentioned inner product on the space of operators.
Note that on the qubit space, such operators are of the form \(\overrightarrow{\sigma} \cdot\widehat{n}\), with \(|\widehat{n}|=1\).

Then,
\begin{equation}
\begin{split}
\sum\limits_i \bra{i}A^2 \ket{i} & =\sum\limits_{i=1}^d \sum \limits_{\mu,\nu=0}^{d^2-1} a_{\mu}a_{\nu} \bra{i}\mathbb{H}_{\mu} \mathbb{H}_{\nu}\ket{i}\\
& =\sum\limits_{\mu,\nu}^{d^2-1} a_{\mu}a_{\nu} \delta_{\mu \nu} \\ 
& =\sum\limits_{\mu} a_\mu^2 \\
& = 1 .
\end{split}
\end{equation}  
%
%
%
%
Let us now set $\ket{\chi_j}$ and $\ket{\phi_i}$ as
\begin{equation} 
 \ket{\chi_j}=A\ket{j}  \quad  \mbox{and}    \quad  \ket{\phi_i}=\ket{i} \bra{i}A\ket{i}.
\end{equation}
So, for normalized $A$, i.e., for $A$ with unit norm, we have  
\begin{equation}
\sum\limits_j \parallel \chi_j \parallel^2 = \sum\limits_j \bra{j} A^2 \ket{j} =1 , 
\end{equation}
and from
\begin{equation}
\begin{split}
\sum\limits_i \bigtriangleup A_i^2 & =\sum\limits_i \bra{i} A^2\ket{i} -\sum\limits_i {\bra{i}A\ket{i}}^2 \\
& =1-\sum\limits_i {\bra{i}A\ket{i}}^2 \\
&  \geqslant 0 , 
\end{split}
\end{equation} 
we get
\begin{equation}
\sum\limits_i \parallel\phi_i\parallel^2 =\sum\limits_i {\bra{i}A\ket{i}}^2 \leqslant 1 . 
\end{equation}
So,
\begin{equation}
\begin{split}
\sum\limits_{i,j} \big|A_{ji} A_{ii}\big|&=\sum\limits_{i,j}\big|\braket{\chi_j|\phi_i}\big|\\
&  \leqslant \sum\limits_{i,j}\parallel\chi_j\parallel \parallel\phi_i\parallel  \\
& \leqslant 1 .
\end{split}
\end{equation}
Now, let us estimate the modulus of the expression derived in (\ref{ekhane-amar-maTidesh}):
\begin{eqnarray}
&&\Big|\text{Im}\sum_{i,j}p_{ij}A_{ji}A_{ii}\Big|\nonumber\\
&=&\Big|\text{Im}\big[\sum_{i \ne j}p_{ij}A_{ji}A_{ii}+\sum_i p_{ii}A_{ii}^2\big]\Big|\nonumber\\
&\leqslant&\Big|\text{Im}\sum_{i \ne j}p_{ij}A_{ji}A_{ii}\Big|\nonumber\\
&\leqslant&\Big|\sum_{i \ne j}p_{ij}A_{ji}A_{ii}\Big|\nonumber\\
&\leqslant&\sum_{i \ne j}\Big|p_{ij}\Big|\Big|A_{ji}A_{ii}\Big|\nonumber\\
&\leqslant&\sum_{i \ne j}\Big|p_{ij}\Big|=C(\rho),
\end{eqnarray}
where $C(\rho)$ measures the quantum coherence, in the reference basis, of the $d$-dimensional quantum state $\rho$ (see Eq.~(\ref{footboard-jhule-jawa}). 

Therefore, for an arbitrary \(d\)-dimensional state $\rho$ and an arbitrary traceless observable of unit norm on the same system, we 
have the relation
\begin{equation}
\label{nadu-mallik}
\frac{1}{2} \Big|\text{tr}\big([A,A^D]\rho\big)\Big| \leqslant C(\rho),
\end{equation}
the qubit version of which was already derived in~(\ref{ektara}). 

We have until now been considering traceless observables, \(A\), with unit norm. An arbitrary Hermitian operator, \(\tilde{A}\), can be written as \(\tilde{A} = a_0 I + aA\), where \(a_0\) and \(a\) are real numbers. Now \(\tilde{A}^D = a_0I + aA^D\), and \([\tilde{A},\tilde{A}^D] = a^2 [A,A^D]\). Therefore, the relation~(\ref{nadu-mallik}) remains valid for arbitrary \(a_0\). It is also valid for arbitrary \(|a|\leq 1\), but becomes weaker for \(|a|<1\). For \(|a|>1\), the relation may get violated, in general.  
\\

\noindent \textbf{Witness of quantum coherence.}
Although the derived relation provides a quantitative estimate of quantum coherence of an arbitrary quantum state, it of course provides a witness for the same. Given a quantum system in a state \(\rho\), we can prove that it has a nonzero quantum coherence in a given reference basis, i.e., witness its quantum coherence, by somehow showing that the commutator \([A, A^D]\) has a nonzero value in the state \(\rho\). For estimating or witnessing quantum coherence, by using the proposed method, we need to begin with choosing an observable \(A\) that is traceless and has unit norm. Note that since \(A\) and \(A^D\) are both Hermitian,  \(i[A, A^D]\) is also Hermitian, and 
\(|\langle i[A, A^D] \rangle| = |\langle [A, A^D] \rangle|\). Therefore, if we are able to find the average of the observable \(i[A, A^D]\), we will have an estimate of the quantum coherence of the state under consideration, or its witness.

\section{The relation for arbitrary mixed states using  convex-roof measure of coherence}
\label{chharpokader-khulechhe-borat}


In the preceding section, we have derived a relation of the noncommutativity of an observable and its incoherent part
with quantum coherence, for an arbitrary \(d\)-dimensional quantum  state, irrespective of whether it is pure or mixed. 
Here we want to check whether the same inequality relation holds also for a convex-roof measure of quantum coherence. 

Consider an arbitrary quantum state \(\rho\) on the \(d\)-dimensional Hilbert space \(\mathbb{C}^d\). Consider now the quantum coherence of this state in the reference basis, as quantified by the convex roof-based \(l_1\)-norm of quantum coherence, as defined in Eq. (\ref{neche-neche-ai-ma-shyama}).
Suppose now that the quantum coherence 
\(\tilde{C}(\rho)\) of \(\rho\) is attained as a limit over a sequence of decompositions 
\begin{equation}
\sum_ip_i^{(n)}|\psi_i^{(n)}\rangle \langle \psi_i^{(n)}|
\end{equation}
of \(\rho\), with \(n\) being the running index of the sequence, so that 
\begin{equation}
\tilde{C}(\rho)= \lim_{n\to\infty}\sum_ip_i^{(n)} C(|\psi_i^{(n)}\rangle).
\end{equation}
For any member of the above sequence of decompositions, we 
have 
\begin{eqnarray}
\Big|\langle[A,A^D]\rangle_\rho\Big| &\equiv& \Big|\mbox{tr}([A,A^D] \rho)\Big| \nonumber \\
&=& \Big|\sum_i p_i \langle \psi_i^{(n)}|[A,A^D]|\psi_i^{(n)}\rangle\Big| \nonumber \\
&\leqslant & \sum_i p_i \Big|\langle \psi_i^{(n)}|[A,A^D]\Big|\psi_i^{(n)}\rangle\Big|,
\end{eqnarray}
wherein we invoke the relation in inequality (\ref{nadu-mallik}), but restricted to \emph{pure states}, to get
\begin{equation}
\Big|\frac{1}{2}\langle[A,A^D]\rangle_\rho\Big| \leqslant 
\sum_i p_i^{(n)} C(|\psi_i^{(n)}\rangle).
\end{equation}
Considering the limit as \(n\to\infty\), we get the desired relation
\begin{equation}
\frac{1}{2}\Big|\langle[A,A^D]\rangle_\rho\Big| \leqslant 
\tilde{C}(\rho),
\end{equation}
for an arbitrary quantum state \(\rho\) on a \(d\)-dimensional Hilbert space, and for an arbitrary traceless observable \(A\)  of unit norm on the same space. 

\section{Discussion}
\label{phurtir-pran-gaRer-maTh}

The 
concepts of noncommutativity and the superposition principle are
two basic features that lie at the heart of a large section of  modern science. In this
paper, we have established a lower bound on the quantum coherence of an
arbitrary quantum state of a quantum system
in arbitrary dimension by utilizing a noncommutativity estimate
of an arbitrary observable of sub-unit norm for that system, where the said estimate is the commutator of the observable and its incoherent part. 


We wish to look at the relation from two perspectives. On the fundamental side, this relation gives us a potential bridge between the concept of noncommutativity within the realm of quantum uncertainty relations and the concept of quantum coherence. On the other hand, the relation provides us with a method to estimate lower bounds of quantum coherence without existing witnesses \cite{eiTa-coherence-witness} or quantum state tomography 
\cite{eiTa-coherence-tomo}. Precisely, we have to measure
the commutator of any observable of sub-unit norm and its incoherent part.
The measurement has to be performed on the quantum state whose quantum coherence we wish to estimate.
And the incoherent part of the observable has to be considered with respect to the reference basis of our quantum coherence measure. 
One-half of the modulus of this value provides a lower bound of the quantum coherence of our state in the reference basis.
We believe that the relation can be
easily tested and put to use with existing experimental quantum information 
setups.

It is true that computing quantum coherence of a quantum state with respect to the eigenbasis of the available observable will not be useful to obtain an estimate of the coherence. We however envisage a situation where one can measure an observable that is not diagonal in the reference basis in which the quantum coherence is sought. As an example, in case the system considered is the polarization degree of freedom of a photon, we can try to measure the observable corresponding to \(i[A, A^D]\), for which the observable \(A\) measures the circular polarization, while the quantum coherence that is sought, is with respect to the basis of horizontal and vertical polarizations.

We intend in future to study how close the quantity \(\frac{1}{2}|\langle [A, A^D] \rangle|\) can be to the quantum coherence of the corresponding state, for an arbitrary choice of \(A\). We will also attempt to derive similar relations for other measures of quantum coherence
of quantum states and noncommutativity estimates of quantum observables.

\begin{acknowledgments}
The research of T.R. was supported in part by the INFOSYS scholarship for senior students. We acknowledge partial support from the Department of Science and Technology, Government of India through the QuEST  grant (grant numbers DST/ICPS/QUST/Theme-1/2019/117 and DST/ICPS/QUST/Theme-3/2019/120).
\end{acknowledgments}


\end{document}